\let\ifarxiv=\iftrue     
{}
{}

\newif\ifpublic\publictrue

\ifarxiv

\documentclass[11pt,a4paper]{article}

\pdfoutput=1
 \usepackage{overpic}

\DeclareMathAlphabet{\mathpzc}{OT1}{pzc}{m}{it} 
\usepackage[mathscr]{euscript}

\usepackage[usenames,dvipsnames]{xcolor}
\usepackage{amsmath,amssymb}
\ifpublic\else\usepackage{showkeys}\fi
\usepackage{graphicx}
\usepackage[bookmarks=true,hyperfigures=true]{hyperref}
\usepackage{sectsty}

\usepackage{mathrsfs}

\sectionfont{\Large}
\subsectionfont{\normalsize\bfseries}
\hypersetup{pdftitle={Charged structure constants from modularity},    
	pdfauthor={ddp}, 
	colorlinks=true, 
	linkcolor=blue, 
	citecolor=green!39!black!,
	filecolor=OliveGreen, 
	urlcolor=RoyalBlue}

\def\showkeysrefformat#1{{\normalfont\tiny\ttfamily#1}}
\makeatletter
\def\SK@@ref#1>#2\SK@{%
 {\@inlabelfalse\leavevmode\vbox to\z@{%
 \vss\SK@refcolor\rlap{\vrule\raise .75em%
  \hbox{\showkeysrefformat{#2}}}}}}
\makeatother

\usepackage[noadjust]{cite}

\usepackage[a4paper,text={162mm,247mm},centering]{geometry}

\allowdisplaybreaks[3]

\usepackage[font=small,labelfont=bf,width=0.85\textwidth]{caption}

\fi

\ifarxiv\else

\iftrue
\PassOptionsToClass{showpacs,showkeys,preprintnumbers}{revtex4-1}
\PassOptionsToClass{superscriptaddress}{revtex4-1}
\fi
\documentclass[10pt,letterpaper,twocolumn,amsmath,amssymb,final,aps,pra]{revtex4-1}

\usepackage{graphicx}
\usepackage[bookmarks=true,hyperfigures=true]{hyperref}
\usepackage[usenames, dvipsnames]{color}

\makeatletter
\let\o@a@f\@author@finish
\def\@author@finish{\o@a@f%
\let\@authors\empty\def\AF@opr##1{}\def\CO@opr##1##2##3{}%
\def\AU@opr##1##2##3{\ifx\@authors\@empty\toks@\expandafter{##2}%
  \else\toks@\expandafter{\@authors, ##2}\fi\edef\@authors{\the\toks@}}
\@AAC@list}
\makeatother

\fi

\def\blue#1{\textcolor{black}{#1}}

\expandafter\def\expandafter\bfseries\expandafter{\bfseries\ifmmode\else\boldmath\fi}
\expandafter\def\expandafter\mdseries\expandafter{\mdseries\ifmmode\else\unboldmath\fi}
\expandafter\def\expandafter\normalfont\expandafter{\normalfont\ifmmode\else\unboldmath\fi}

\usepackage{setspace}

\let\barefrac=\frac
\renewcommand{\frac}[2]{\mathinner{\barefrac{#1}{#2}}}

\let\baresqrt=\sqrt
\makeatletter
\renewcommand{\sqrt}{\@ifnextchar[\@sqrt@space@a\@sqrt@space@b}
\def\@sqrt@space@a[#1]#2{\mathinner{\mathchoice{\mkern-3mu}{\mkern-3mu}{}{}\baresqrt[#1]{#2}}}
\def\@sqrt@space@b#1{\mathinner{\mathchoice{\mkern-3mu}{\mkern-3mu}{}{}\baresqrt{#1}}}
\makeatother

\makeatletter
\let\per@dot@old=\.
\def\.{\ifmmode\def\per@dot@sel{\mkern3mu}\else\def\per@dot@sel{\per@dot@old}\fi\per@dot@sel}
\makeatother

\newcommand{\vfrac}[2]{\ifmmode\mathinner{\textstyle^{#1}\!/\!_{#2}}\else$^{#1}\!/\!_{#2}$\fi}

\makeatletter\let\Re\@undefined\let\Im\@undefined\makeatother
\DeclareMathOperator{\Re}{Re}
\DeclareMathOperator{\Im}{Im}
\DeclareMathOperator{\tr}{tr}

\newcommand{\nn}{\nonumber}

\def\[{\begin{equation}}
\def\]{\end{equation}}

\providecommand{\href}[2]{#2}

\makeatletter
\def\mr@ignsp#1 {\ifx\:#1\@empty\else #1\expandafter\mr@ignsp\fi}%
\newcommand{\multiref}[1]{\begingroup
\xdef\mr@no@sparg{\expandafter\mr@ignsp#1 \: }%
\def\mr@comma{}%
\@for\mr@refs:=\mr@no@sparg\do{\mr@comma\def\mr@comma{,}\ref{\mr@refs}}%
\endgroup}
\renewcommand{\eqref}[1]{(\multiref{#1})}
\makeatother

\makeatletter
\newcommand{\namedref}[2]{\hyperref[#2]{#1~\ref*{#2}}}
\newcommand{\secref}{\@ifstar{\namedref{Section}}{\namedref{Sec.}}}
\newcommand{\appref}{\@ifstar{\namedref{Appendix}}{\namedref{App.}}}
\newcommand{\tabref}{\@ifstar{\namedref{Table}}{\namedref{Tab.}}}
\newcommand{\figref}{\@ifstar{\namedref{Figure}}{\namedref{Fig.}}}
\makeatother

\let\oldbib=\thebibliography
\def\thebibliography{\phantomsection\addcontentsline{toc}{section}{\refname}\oldbib}

\let\oldtoc=\tableofcontents
\def\tableofcontents{\phantomsection\addcontentsline{toc}{section}{\contentsname}\oldtoc}

\newcommand{\uone}{\mathfrak{u}(1)}

\providecommand{\hypersetup}[1]{}

\hypersetup{plainpages=false}
\hypersetup{pdfpagemode=UseNone}
\hypersetup{bookmarksnumbered=true}
\hypersetup{pdfstartview=FitH}

\makeatletter
\let\@keywords\@empty
\let\@subject\@empty
\providecommand{\keywords}[1]{\gdef\@keywords{#1}}
\providecommand{\subject}[1]{\gdef\@subject{#1}}
\def\thetitle{\@title}
\def\theauthor{\@author}
\def\thesubject{\@subject}
\def\thedate{\@date}
\def\thekeywords{\@keywords}
\makeatother
\AtBeginDocument{
\hypersetup{pdftitle={\thetitle}}%
\hypersetup{pdfauthor={\theauthor}}%
\hypersetup{pdfsubject={\thesubject}}%
\hypersetup{pdfkeywords={\thekeywords}}%
}

\ifpublic
\newcommand{\remark}[2][]{\ignorespaces}
\else
\usepackage{ifpdf}
\ifpdf\else\RequirePackage[active]{srcltx}\fi
\RequirePackage{color}

\newcommand{\remark}[2][]{{\normalfont\sffamily\hspace{1ex}\def\tmparga{#1}%
  \def\tmpargb{MR}\ifx\tmparga\tmpargb\color{magenta}\fi%
  \def\tmpargb{NB}\ifx\tmparga\tmpargb\color{blue}\fi%
  \def\tmpargb{AG}\ifx\tmparga\tmpargb\color{Orange}\fi%
  \def\tmpargb{Refs}\ifx\tmparga\tmpargb\color{Orchid}\fi%
  \def\tmpargb{}\ifx\tmparga\tmpargb\color{red}\fi%
  \def\tmpargb{}\ifx\tmparga\tmpargb\else \textbf{#1:} \fi%
  #2\hspace{1ex}}}
\fi

\newcommand{\remarkref}[1][]{{\def\tmparga{#1}\def\tmpargb{}%
  \ifx\tmparga\tmpargb\remark[Refs]{needed}\else\remark[Refs]{#1}\fi}}

\def\nd{{ \vphantom{\dagger}}}

\def\cN{\mathcal{N}}

\def\ie{{\it i.e.~}}
\def\eg{{\it e.g.~}}

\def\bra#1{{\langle}#1|}
\def\ket#1{|#1\rangle}
\def\vev#1{\langle #1 \rangle}

\def\nn{\nonumber}
\def\pd{\partial}
\def\Re{R\'{e}nyi }
\def\l1{{{1-loop}}}

\def\by{{\bar{y}}}

\def\n1{\Bigg|_{n=1}}

\def\n{{(n)}}
\def\tr{{Tr}}

\def\cN{\mathcal{N}}

\def\tr{\text{Tr}}

\def\blue#1{\textcolor{blue}{#1}}
\def\cZ{\mathcal{Z}}

\def\cL{\mathcal{L}}
\def\la{\langle}
\def\ra{\rangle}

\def\cF{\mathcal{F}}
\def\opr{\mathbb{O}}
\def\bq{\bar{q}}

\def\state{{\mathrm \Upsilon}}

\def\coeft{{\mathscr C}}

\def\q{{\mathpzc{q}}}
\def\bq{\bar{\mathpzc{q}}}

\def\ausricht{\begin{aligned}}
\def\endeausricht{\end{aligned}}

\begin{document}
\title{\LARGE Charged structure constants from modularity}
\long\def\theabstract{
We derive a universal formula for the average {heavy-heavy-light} structure constants for $2d$ CFTs with non-vanishing $\uone$ charge. The derivation utilizes the modular properties of one-point functions  on the torus. Refinements in   $\mathcal{N}=2$ SCFTs, show that the resulting Cardy-like formula for the structure constants has precisely the same shifts in the central charge as that of the thermodynamic entropy found earlier. This analysis generalizes the recent results by Kraus and Maloney for CFTs with an additional global $\uone$ symmetry \cite{Kraus:2016nwo}.  Our results at large central charge are also shown to match with computations from the holographic dual, which suggest that the averaged CFT three-point coefficient also serves as a useful probe of detecting black hole hair. 
}

\ifarxiv\else
\author{Diptarka Das}
\email{diptarka@physics.uscd.edu}
\affiliation{Department of Physics,\\ University of California San Diego,\\
	La Jolla, CA 92093, USA}
\author{Shouvik Datta}
\email{shouvik@itp.phys.ethz.ch}
\affiliation{Institut f\"ur Theoretische Physik, 
Eidgen\"ossische Technische Hochschule Z\"urich, 
Wolfgang-Pauli-Strasse 27, 8093 Z\"urich, Switzerland}

\author{Sridip Pal}
\email{srpal@ucsd.edu}
\affiliation{Department of Physics,\\ University of California San Diego,\\
	La Jolla, CA 92093, USA}

\begin{abstract}
\theabstract
\end{abstract}

\maketitle
\fi

\ifarxiv
\pdfbookmark[1]{Title Page}{title}
\thispagestyle{empty}


\vspace*{0cm}
\begin{center}%
\begingroup\Large\bfseries\thetitle\par
\endgroup
\vspace{2cm}

\vspace{.5cm}%

\begingroup\scshape
Diptarka Das\textsuperscript{$\tau$}, 
Shouvik Datta\textsuperscript{$-1/\tau$} and 
Sridip Pal\textsuperscript{$\tau$}
\endgroup
\vspace{5mm}

\textit{\textsuperscript{$\tau$}Department of Physics,\\ University of California San Diego,\\
	9500 Gilman Drive, \\
	La Jolla, CA 92093, USA.}
\vspace{0.1cm}

\begingroup\ttfamily\small
\verb+{+didas,srpal\verb+}+@ucsd.edu\par
\endgroup
\vspace{5mm}

\textit{\textsuperscript{$-1/\tau$}Institut f\"ur Theoretische Physik,\\
Eidgen\"ossische Technische Hochschule  Z\"urich,\\
Wolfgang-Pauli-Strasse 27, \\
{}\ \ 8093 Z\"urich, Switzerland.}
\vspace{0.1cm}

\begingroup\ttfamily\small
shdatta@ethz.ch\par
\endgroup
\vspace{1.9cm}
\vfill

\textbf{Abstract}\vspace{5mm}

\begin{minipage}{15.5cm}
\theabstract
\end{minipage}

\vspace*{4cm}

\end{center}

\newpage
\fi



\section{Introduction}
\label{sec:intro}

The most important pieces of data of a conformal field theory   are its central charge, the spectrum of primaries and three point coefficients.  Crossing symmetry of 4-point functions  imposes severe restrictions  on this data.    This circle of ideas form the core of the conformal bootstrap programme,  which has seen a lot of development over the recent years \cite{ElShowk:2012ht,Rattazzi:2008pe,Rychkov:2015naa}.  In (1+1) dimensions, additional constraints can be obtained from modular invariance on the torus. This has proven to be useful in deriving a number of strong and universal results  \cite{Hellerman:2009bu,Friedan:2013cba,Hartman:2014oaa,Collier:2016cls,Collier:2017shs,Cho:2017fzo,Ribault:2014hia}. Amongst its many uses, this plays an important role in the context of the AdS/CFT correspondence. It can enable us to derive some universal features which are relevant for holographic CFTs. One can thereby hope to understand holography and gain better insights into the information paradox.

%
%

Not only does modular invariance impose constraints on 2$d$ CFTs, it also relates low energy data to high energy data. The celebrated exemplar  of the holographic manifestation of a CFT is that of the Cardy formula \cite{cardy1986operator,Carlip:1998qw}. This universal formula captures the black hole entropy, given by the area of the horizon.
Since the advent of the AdS/CFT correspondence, a number of similar relations have been reaped which display the interplay between various quantities of the CFT and geometric objects in the bulk. The noteworthy laurels in this list include : entanglement entropy (related to minimal surfaces in the bulk) \cite{Ryu:2006bv} and conformal blocks (related to geodesics/Witten diagrams) \cite{Fitzpatrick:2014vua,Hijano:2015qja,Hijano:2015zsa,Hijano:2015rla}, to name a  few.

A very recent development along these lines, has been to derive the  \textit{averaged} three point coefficient, \ie of two heavy and one light operator, using modular properties of one-point functions on the torus \cite{Kraus:2016nwo}. In a sense, this goes beyond the Cardy formula, since the partition function itself is given by the one-point function of the identity operator. 
 In this paper, we generalize the analysis of \cite{Kraus:2016nwo} to the case when the CFT has a global $\uone$ symmetry and a non-zero chemical potential for the same. That is, we are interested in theories whose partition function has a $\uone$ grading. In other words, this is the grand canonical partition function, given by
\begin{align}
\cZ(\tau, \nu)=\tr\left[\q^{L_0-c/24}y^{J_0}\bar \q^{\bar L_0-c/24}\by^{\bar J_0}\right]. \nn 
\end{align}
Owing to the presence of chemical potentials, such partition functions transform as weak Jacobi forms. We shall make use of these modular transformation properties  to attain an analogous formula for the three-point coefficients.

 Most of our results cover the large class of CFTs which contain Virasoro$\,\times \, \uone$ Kac-Moody as their chiral algebra. These include, most importantly $\cN=2$ SCFTs and also theories with $\mathcal{W}_{1+\infty}$ symmetry. We shall make very few assumputions on the spectrum; however, our results are valid for any value of the central charge. We also provide further refinements of the formula for three-point coefficients of three primaries in $\cN=2$ SCFTs. The resulting expression displays precise shifts in the central charge which has been observed earlier in the exact formula for thermal entropy  \cite{Benjamin:2016aww}.


The three-point coefficients also admit a description in terms of geodesics in the AdS dual. The dual background for our CFT setup is that of the BTZ black hole charged with $\uone$ hair. This is a solution of the Einstein-Chern-Simons theory. We shall be able to reproduce the structure constants by evaluating the length of an appropriate geodesic network. Furthermore, since the black hole is charged, we shall see that bulk Wilson loops are also necessary in order to reproduce CFT result in its entirety. 
It has been discussed in \cite{Kraus:2016nwo} that these CFT results combined with the matching from holographic calculations in the black hole geometry hints at the notion of the black hole geometry being an averaged version of heavy microstates. The result of the present  work therefore extends this picture to case when the black hole has additional charged hairs.

\subsection*{Summary of main results}
\label{sec:result}
For the convenience of the readers, we provide the main findings of our analysis  here. We are interested in the three-point coefficient of an uncharged light operator $\opr$ with two other heavy operators (denoted by $\state$) with conformal dimension $\Delta$ and charge $(Q,\bar Q)$. 
The  mean value of this quantity shall be denoted by $\mathscr{C}_{{\state^\dagger \opr^\nd \state^\nd}}$ and is defined by the ratio of the spectral density weighted with the 3-point coefficient and the spectral density itself.   In the limit $\Delta \rightarrow \infty$, we have
\begin{align}\label{main-result}
{\mathscr C}_{{\state^\dagger \opr^\nd \state^\nd}}^{\text{}}  &\approx N_{\opr} \,  
c_{\chi^\dagger \opr^\nd \chi^\nd}
\bigg( \Delta - \tfrac{Q^2 + \bar{Q}^2}{2k} -\frac{c}{12} \bigg)^{\Delta_\opr /2}    \nn \\
 &{} \exp\bigg[ -\tfrac{\pi c}{3} \sqrt{ \tfrac{12}{c}\bigg( \Delta - \tfrac{Q^2 + \bar{Q}^2}{2k}\bigg) -1 } \bigg\{ 1 - \sqrt{ 1 - \tfrac{12}{c}\bigg( \Delta_{\chi} - \tfrac{q_\chi ^2 + \bar{q}_\chi^2}{2k}\bigg) } \bigg\} \bigg]e^{-\tfrac{2\pi i }{k} ( q_\chi Q - \bar{q}_\chi \bar{Q} )}. \nn
\end{align}
Here, $\chi$ is the lightest charged primary -- with conformal dimension $\Delta_\chi$ and $\uone$ charge $(q_\chi,\bar q_\chi)$. $N_\opr$ is a constant independent of $(\Delta,Q,\bar Q)$. Here, $k$ is the level of the $\uone$ Kac-Moody algebra. {In case of the $\mathcal N =2$ superconformal algebra, the standard convention relates the level to the central charge by $k=c/3$}. 
It is useful to contrast ${\mathscr C}_{{\state^\dagger \opr^\nd \state^\nd}}^{\text{}} $ with the uncharged result of  \cite{Kraus:2016nwo}  
\begin{align} \nn 
{ C}_{\state^\dagger \opr  \state} &\approx N_\opr \,  c_{\chi^\dagger  \opr \chi} \, 
\bigg( \Delta -\frac{c}{12} \bigg)^{\Delta_\opr /2}  
 \exp\bigg[ -\frac{\pi c}{3} \sqrt{ \frac{12\Delta}{c}  -1 } \bigg\{ 1 - \sqrt{ 1 - \frac{12\Delta_\chi}{c} } \bigg\} \bigg].
\end{align}
We notice that, in addition to the spectral flow shifts of the conformal dimensions of the charged primaries,   the charged structure constants also possess additional phase which is  of the form $e^{ -2\pi i( q_\chi Q - \bar{q}_\chi \bar{Q} )/k }$. 
The `holographic'   large $c$ limit ($c \gg \Delta_\chi, c\gg \Delta_O$) of this formula shall be reproduced from AdS$_3$ gravity; the dual background being that of a charged BTZ black hole\footnote{In a related context, heavy light conformal blocks in presence of $\uone$ charge has been studied in \cite{Fitzpatrick:2015zha}.}. 


It is also possible to make further refinements to the above formula, for the case of three-point coefficient of three primaries in  $\cN=2$ SCFTs. The derivation uses asymptotic properties of $\cN=2$ torus blocks. The resultant expression for the average three point function has the central charge shifted to $c \mapsto c-3$. This precise shift has been observed earlier in \cite{Benjamin:2016aww} for the Cardy formula for the entropy in $\cN=2$ theories and  {can be holographically understood as one-loop renormalization of the effective central charge due to dressing caused by gravitons, gravitini and the spin-1 gauge field.} This is analogous to the shift to the shift $c\mapsto c-1$, for the non-supersymmetric case, derived in \cite{Kraus:2016nwo}.


\subsection*{Outline }

This paper is organised as follows. In \S\ref{sec:modular} we discuss the relevant modular transformation properties of one-point functions in presence of $\uone$ charge. \S\ref{sec:derivation} contains the derivation of the asymptotics of the charged structure constants. We specialize to $\cN=2$ SCFTs in \S\ref{sec:n=2}. The large central charge limit of the CFT results are reproduced from gravity in \S\ref{sec:einstein}. We conclude in \S\ref{sec:conclusions}. Appendices \ref{app:alg} and \ref{app:u1-CS} provide some of the technical details of our analysis. We discuss the relevance of Tauberian theorems which is relevant to our present context in Appendix \ref{Tauber}.

\section{Modular properties of charged CFTs}
\label{sec:modular}

For a 2$d$ CFT with a global abelian symmetry, 
the torus partition function in  the grand canonical ensemble  (\ie for the non-vanishing chemical potential, $\nu$, for the $\uone$ current)  can be written as, 
\begin{equation}
\cZ(\tau, \nu) = \tr \left[\q^{L_0-\frac{c}{24}} \bar{\q}^{ \bar{L}_0 - \frac{c}{24} } y^{J_0}\bar{y}^{\bar{J}_0} \right].
\end{equation}
Here,  the nome and the fugacity are given by  $\q = e^{ 2\pi i \tau } $ and   $y = e^{2\pi i \nu }$ respectively\footnote{It is rather unfortunate that the usual $q$ will be used to denote the $\uone$ charge in due course.}. $J_0$ is the zero-mode of the $\uone$ current.  The modular parameter of the torus given in terms of the inverse temperature $\beta$ and the circumference of the spatial circle, $L$ $$\tau = i \beta/L.$$ 
The modular transformation on $\nu$ takes real $\nu$ to imaginary $\nu$ and vice versa. Therefore, $\nu$ is chosen to be complex $\mu_{R}+ i\mu_{I}$ to keep things general.
This means that the chemical potentials for both the $\uone$ charge ($J_0+\bar J_0$) and the fermion number ($J_0 - \bar J_0$) are turned on. It is known  that under a modular transformation, $\gamma \cdot \tau = \frac{ a\tau + b}{c \tau + d}$, the partition function transforms as \cite{Benjamin:2016fhe,Gaberdiel:2012yb}\footnote{See also Appendix A of \cite{deBoer:2014fra} for an example.}
\begin{equation}\label{parti}
\cZ(\tau,\nu) \ \ \mapsto \ \  \cZ\left(  \frac{ a\tau + b}{c \tau + d},\frac{\nu}{c\tau + d} \right) = \exp\bigg( \frac{ ic\pi k \nu^2}{c\tau + d} - \frac{ic\pi k \bar{\nu}^2}{c\bar{\tau} + d } \bigg) \cZ(\tau,\nu) .
\end{equation}
Here, $k$ is the level of the $\uone$ Kac-Moody algebra. 
We also need to consider how a primary operator, $\opr$,   transforms under modular transformations. 
The operator $\opr$  is located at the complex elliptic coordinates of the torus, $w$ and $\bar{w}$. 
 Under the modular group PSL$(2,\mathbb{Z})$, the elliptic coordinate  transforms as \cite{eichler-zagier}
\begin{align}
\gamma \cdot w = \frac{w}{c\tau +d}  . 
\end{align}
Consider a primary operator $\opr$ obeying $[L_{n>0},\opr]=0$ and $[J_{n\geq 0}, \opr]=0$. Under modular transformations, it transforms as 
\begin{align}\label{op-trans}
\ausricht
\opr (w,\bar w) |_{\tau} &= \left[\frac{\pd (\gamma \cdot w)}{\pd w}\right]^{h}\left[\frac{\pd (\gamma \cdot \bar w)}{\pd \bar w}\right]^{\bar h} \opr (\gamma \cdot w,\gamma \cdot \bar w)   \\ & = (c\tau+d)^{-h} (c\bar \tau+d)^{-\bar h}\, \opr (\gamma \cdot w,\gamma \cdot \bar w)\big|_{a\tau + b \over c\tau +d } .
\endeausricht 
\end{align}
The one-point function of the operator $\opr$ on the torus is given by 
\begin{align}\label{one-point-def}
\la \opr (w,\bar w)\ra_{\tau,\nu} \equiv \tr \left[ \opr(w, \bar w) \q^{L_0-c/24}y^{ J_0}\bar \q^{\bar L_0-c/24}\bar y^{\bar J_0}\right].
\end{align}
Combining \eqref{parti} and \eqref{op-trans}, we are led to the following transformation property for one-point function of primaries on the torus (see also \cite{Gaberdiel:2009vs}) 
\begin{align}\label{one-point-trans}
\vev{\opr (\gamma \cdot w,\gamma \cdot \bar w)}_{\tau,\nu} = (c\tau+d)^{ h} (c\bar \tau+d)^{ \bar h} \exp\bigg( \frac{ ic\pi k \nu^2}{c\tau + d} - \frac{ic\pi k \bar{\nu}^2}{c\bar{\tau} + d } \bigg) \vev{\opr (w,\bar w) }_{{a\tau + b \over c\tau +d} ,{\nu \over c\tau +d }}.
\end{align}
This property will play a key role in determining the asymptotics of the charged structure constants. Note that, the one-point function does not have position dependence, owing to translation symmetry on the torus. Moreover, the chiral half of the transformation formula \eqref{one-point-trans} is that of a \textit{weakly holomorphic Jacobi form} of weight $h$ and index $k$ \cite{DMZ}.  

\section{Derivation of the asymptotic formula }
\label{sec:derivation}
We start with the one-point function of a neutral primary on the torus with modular parameter, $\tau = i \beta/L$ and chemical potential $\nu = \mu_{R}+ i \mu_{I}$. The trace in equation \eqref{one-point-def} can be rewritten as sum of states as
\begin{align}\label{GCE-def}
\vev{\opr}\ \   
&= \ \ \sum\limits_{\alpha} \bra{\state_\alpha}\opr \ket{\state_\alpha} \,  \q^{\Delta_\alpha-\frac{c}{24}} \, \bar{\q}^{ \Delta_\alpha  - \frac{c}{24} } y^{Q_\alpha}\bar{y}^{\bar{Q}_\alpha}. 
\end{align}
where, the state $\ket{\state_\alpha}$ has the following eigenvalues 
\begin{align}\label{state-def}
(L_0+\bar{L}_0) \ket{\state_\alpha} = \Delta_\alpha \ket{\state_\alpha}, \qquad J_0 \ket{\state_\alpha} = Q_\alpha\ket{\state_\alpha}, \qquad \bar{J}_0 \ket{\state_\alpha} = \bar{Q}_\alpha\ket{\state_\alpha}. 
\end{align}
It is worthwhile to mention here that we are working with the \textit{unnormalized} one-point function; in the sense that we do not divide out by the partition function. This convention has its advantages; namely, the one-point function of the identity operator is then the partition function itself. 
 In the low temperature regime, $\beta/L \rightarrow \infty$, using the $\q$-expansion of the sum above, we can  write 
\begin{equation}\label{lowexp}
\vev{ \opr }_{\text{low-temp}} \ \simeq \ 
 \langle \chi|\opr|\chi\rangle e^{ -2\pi \beta/L ( \Delta_{\chi} - \frac{c}{12} )} e^{ 2\pi i \mu_{R}q_- }e^{-2\pi \mu_{I}q_+}+ \cdots.
\end{equation}
where, $\chi$ is the lightest primary with $c_{\chi^\dagger\opr^\nd \chi^\nd} \neq 0$ \footnote{We assume that $\chi$ is non-degenerate  
 and  $c_{\chi^\dagger O^\nd \chi^\nd}$ is not exponential in the conformal weight. 
 } and we have defined $q_\pm = q_\chi\pm \bar{q}_\chi$ in terms of the charges of $\chi$. The following analysis is also valid even when $\chi$ is neutral, \ie $q_\chi = \bar{q}_\chi =0$. However, more interestingly one can indeed find theories in which the lightest state is charged. 
 For example in $\cN=2$ super-Virasoro minimal models, the lightest state, $\chi$, has the quantum numbers
 \begin{align}
 h_{\chi} = \frac{1}{2(k+2)}, \qquad q_{\chi} = \pm \frac{1}{k+2}. 
 \end{align}
In this specific case, $\chi$ is additionally BPS, satisfying $h=\pm\, q/2$.

The light operators $\chi$ and $\chi^\dagger$ carry opposite $\uone$ charges. The presence of the terms in the OPE
\begin{align}\label{fusion}
\chi^{\dagger} \chi \sim \mathbb{I} + \mathbb{O} + \cdots .
\end{align}
are consistent with charge conservation,
\ie 
 the operators appearing on the right hand side are neutral under the $\uone$. We also require that $\chi$ falls within the bound of   \cite{Benjamin:2016fhe}.

 From the low temperature expansion \eqref{lowexp}, we can get the expansion in the high temperature regime by performing the S-modular transformation. This is tantamount to choosing $a = d = 0; c = -b = 1$ \footnote{This can also  be achieved via $b=-c=1$. However, this results in a  different representation of $\opr$ under the modular group, PSL$(2,\mathbb{Z})$.}, which  takes $\tau \rightarrow -1/\tau $ and $\nu \rightarrow \nu/\tau$. Hence,  using equations \eqref{one-point-trans} and \eqref{lowexp} we have,
\begin{equation}\label{highT}
\vev{\opr }_{\text{high-temp}} \simeq 
i^{-S_O }    \langle \chi|\opr|\chi\rangle e^{ - \frac{ 2\pi L k}{\beta} ( \mu_R^2 - \mu_I^2 ) } \bigg( \frac{ L} { \beta} \bigg)^{\Delta_\opr} e^{ -\frac{2\pi L}{\beta} ( \Delta_\chi - \frac{c}{12} ) } e^{ \frac{ 2 \pi  L q_+ }{\beta} \mu_R }e^{\frac{ 2 \pi i  L q_- }{\beta} \mu_I }.
\end{equation} 
Let us now rewrite the summation over states in the grand canonical expectation value \eqref{GCE-def}  as an integral 
\begin{align}
& \vev{\opr }_{} 
 = \int d\Delta dQ_+ dQ_-\, \, T_\opr( \Delta,Q_\pm ) \bigg(\frac{2\pi}{L}\bigg)^{\Delta_{\opr}} i^{-S_{\opr}} e^{-\frac{2\pi \beta}{L} ( \Delta - \frac{c}{12} ) } e^{ 2\pi i \mu_R Q_-} e^{ -2\pi \mu_I Q_+ },\label{tracedef}
\end{align}
where we have used $Q_{\pm} = Q\pm \bar{Q}$ which are eigenvalues of the operators $J_0$ and $\bar{J}_0$. $T_O(\Delta, Q_\pm)$ is the `weighted spectral density'   -- \ie the density of states weighted by the three-point coefficient -- and is defined as 
\begin{eqnarray}\label{TQ-def}
T_\opr(\Delta,Q_\pm)   &=&  \sum\limits_{\alpha} \, c_{\state_\alpha^\dagger \opr ^\nd \state_\alpha^\nd} \, \delta(\Delta - \Delta_\alpha) \delta( Q_\pm - Q_{\alpha,\pm}) \, ,
\end{eqnarray}
where, $\rho(\Delta,Q_\pm)$ is the density of states at conformal dimension $\Delta$ and charges $Q_\pm$ ({this is the `unweighted spectral density'})
\begin{align}
\rho(\Delta,Q_\pm)   &=  \sum\limits_{\alpha}  \delta(\Delta - \Delta_\alpha) \delta( Q_\pm - Q_{\alpha,\pm})
\end{align}
In the above equations and the ones to follow, $\delta( Q_\pm - Q_{\alpha,\pm})$ refers to the product of the $\delta$-functions $\delta( Q_+ - Q_{\alpha,+})\delta( Q_- - Q_{\alpha,-})$.
The notion of the weighted spectral density also appears in the discussion of OPE convergence in \cite{RychkovOPE}. However, the discussion there deals with the spectral density weighted by the square of the OPE coefficients. 

The weighted spectral density \eqref{TQ-def} also defines the \textit{average three point function}, $\coeft_{\state^\dagger \opr ^\nd \state^\nd}$, the key object we are interested in this paper. 
\begin{align}\label{to}
\coeft_{\state^\dagger \opr ^\nd \state^\nd} \ \equiv \ \frac{T_\opr(\Delta,Q_\pm) }{\rho(\Delta,Q_\pm)  }
\end{align}
The average is taken over operators in the CFT  with the fixed scaling dimension $\Delta$ and $\uone$ charge $Q_\pm$.

The integral \eqref{tracedef} can be inverted to find $T(\Delta,Q_\pm)$. This is similar to procedure followed in \cite{Kraus:2016nwo}. However, now the inversions involve one inverse Fourier transform and two inverse Laplace transforms, (due to the additional chemical potentials present)
\begin{equation}\label{inverse}
T_\opr  ( \Delta,Q_\pm ) = \frac{1}{L} \oint d\beta \int_{-\infty}^\infty d\mu_R\oint d\mu_I \, \, \bigg(\frac{2\pi}{L}\bigg)^{-\Delta_{\opr}} i^{S_{\opr}} \vev{ \opr }
e^{\frac{2\pi \beta}{L} ( \Delta - \frac{c}{12} ) } e^{ -2\pi i \mu_R Q_-} e^{ 2\pi \mu_I Q_+ }.
\end{equation}
We also note that it is clear that in   $\beta/L\rightarrow 0$ regime,  the integral  \eqref{tracedef} will be dominated by large $\Delta$. Since we are interested in the asymptotic behavior of three point coefficients of the type heavy-light-heavy, we should then use the high temperature expansion of the torus grand canonical one point function. This is given by equation \eqref{highT} which we substitute in \eqref{inverse}.  
It turns out that the $\mu_{R}$ and $\mu_{I}$ integrals can be done explicitly and the results are as follows:
\begin{align}\label{gauss}
\ausricht
\int_{-\infty}^\infty d\mu_R \, \, e^{ - \frac{ 2\pi L k}{\beta} \mu_R^2  } e^{ \frac{ 2 \pi  L q_+ }{\beta} \mu_R }e^{ -2\pi i \mu_R Q_-} = \sqrt{\frac{ \beta}{2 L k } } \exp\bigg[ \frac{ \pi L }{2k \beta}q_+^2 - \frac{\pi \beta}{2 L k} Q_-^2 \bigg] e^{ - \frac{\pi i }{k} q_+ Q_-  } ,\\
\frac{1}{ i } \int_{-i \infty}^{i \infty} d\mu_I \, \, e^{  \frac{ 2\pi L k}{\beta} \mu_I^2  } e^{\frac{ 2 \pi \imath L q_- }{\beta} \mu_I } e^{ 2\pi \mu_I Q_+ }=\sqrt{\frac{ \beta}{2 L k } } \exp\bigg[  \frac{ \pi L }{2k \beta}q_-^2 -\frac{ \pi \beta  }{2 L k }Q_+^2\bigg] e^{ - \frac{\pi i }{k} q_- Q_+  }.
\endeausricht
\end{align}
The first integral is a Fourier transform of a Gaussian, which yields another Gaussian. The second integral is a Gaussian integral in itself.
We finally collect the $\beta$ dependent factors from equations \eqref{gauss} and \eqref{inverse}, which leave us with the following integral.
\begin{equation}\label{beta-int}
 \oint d\beta\, \,  \beta^{ 1 - \Delta_\opr} \exp\bigg[ \frac{2\pi \beta}{L} \bigg\{ ( \Delta - \frac{c}{12} ) - \frac{1}{4k} ( Q_+^2 + Q_-^2 ) \bigg\} - \frac{2\pi L}{\beta} \bigg\{ ( \Delta_\chi - \frac{c}{12} ) - \frac{1}{4k} (q_+^2+q_-^2)\bigg\}\bigg].
 \end{equation}
We shall be interested in the high energy asymptotics, \ie limit of large $\Delta$. The above integral can hence be evaluated using the saddle point approximation. The saddle point is located at,
\begin{equation}\label{betas}
\beta_s = L \sqrt{ \frac{ \frac{c}{12} - [\Delta_\chi-\frac{1}{4k} ( q_+^2 + q_-^2 ) ] }{ [\Delta- \frac{1}{4k} ( Q_+^2 + Q_-^2 )] - \frac{c}{12} }} + \frac{L}{4\pi } \frac{ \Delta_\opr -1 }{ [\Delta- \frac{1}{4k}  ( Q_+^2 + Q_-^2 )] - \frac{c}{12} } + \mathrm{O}(\Delta^{-3/2}).
\end{equation}
We note that $\Delta_\opr$ shifts the saddle in the sub-leading order in the $1/\Delta$ expansion. 
This saddle needs to be real and   consistent with the high temperature expansion. 
The relevant condition (to investigate the large  $\Delta$ asymptotics) for the reality of the saddle is
\begin{align}\label{ineq}
\Delta - {Q_+^2 + Q_-^2 \over 4k} > \frac{c}{12}, \qquad  \Delta_\chi  - { q_+^2 + q_-^2 \over 4k} < \frac{c}{12}  .
\end{align}
The above inequalities have a satisfying interpretation in the bulk dual which we shall consider in \S\ref{sec:einstein}. The zero-mass BTZ black hole is given by $c/12$. The first inequality implies that the asymptotic energy regime is above the charged BTZ black hole threshold, whilst the scalar corresponding to the light operator $\chi$ is below the same.\footnote{This suggests the existence of light charged particles in $AdS_3$ in the spirit of the Weak Gravity Conjecture \cite{ArkaniHamed:2006dz, Montero:2016tif}.} $\chi$ can therefore be considered as a perturbative bulk scalar or a massive point particle.


%

We shall now adopt the following definitions for clarity of our expressions 
\begin{align}\label{spectral-flow-inv}
\ausricht
 \Delta_Q \equiv  \Delta -  \frac{1}{4k}  ( Q_+^2 + Q_-^2 ) = \Delta - \frac{Q^2 + \bar{Q}^2}{2k}, \qquad 
  \Delta_{\chi,q} \equiv\Delta_\chi - \frac{1}{4k} ( q_+^2 + q_-^2 ) = \Delta_\chi - \frac{q_\chi^2 + \bar{q}_\chi^2}{2k}.
 \endeausricht
\end{align}
 {These} quantities are none other than the \textit{spectral flow invariants} in the charged CFT. See Appendix \ref{app:alg} for a short review of the relevant details.
Including quadratic fluctuations around the saddle point,  we have, 
$$
 \frac{1}{i} \int_{-i\infty}^{i\infty} d\beta \, \, e^{ f(\beta) } \approx e^{f( \beta_s)} \sqrt{\frac{2\pi}{f''(\beta_s)} }.
$$
The final expression for $T_\opr ( \Delta,Q_\pm) $ then reads 
\begin{align}\label{growth}
\ausricht
  T_\opr( \Delta,Q_\pm) =\ &  \frac{i^{-S_\opr}c_{\chi^\dagger \opr ^\nd \chi^\nd}}{2k}  \frac{1}{\sqrt{2}} \bigg( \frac{c}{12} - \Delta_{\chi,q} \bigg)^{3/4-\Delta_O/2}  \bigg(\Delta_Q - \frac{c}{12}  \bigg)^{\Delta_O/2-5/4}   \\
&{\times}  \exp\bigg\{  4\pi \sqrt{ \frac{c}{12} - \Delta_{\chi,q} }\sqrt{ \Delta_Q - \frac{c}{12} } \bigg\} \exp\bigg\{ -\frac{2 \pi i }{k}  \bigg( q_\chi Q - \bar{q}_\chi\bar{Q} \bigg) \bigg\}. 
\endeausricht
\end{align}
One can also  carry out the derivation for the density of states for the charged case retaining the quadratic saddle fluctuations. This is the same calculation as the above with the operator $\opr$ being the vacuum/identity.  This boils down to setting $\Delta_{\chi,q}=0=\Delta_{\mathbb{I}}= \Delta_{Q} $, $Q=0=q$ and using the normalization convention $c_{\mathbb{III}}=1$. The density of states then reads 
\begin{equation}\label{DoS}
\rho(\Delta,Q_\pm ) = \frac{1}{2k} \bigg( \frac{c}{12} \bigg)^{3/4} \frac{1}{\sqrt{2}} \bigg( \Delta_Q - \frac{c}{12} \bigg)^{-5/4} 
\exp\bigg\{ 4\pi \sqrt{ \frac{c}{12} \bigg( \Delta_Q - \frac{c}{12} \bigg) }\bigg\}. 
\end{equation}
We can get the asymptotics of  the mean charged three point coefficient from \eqref{TQ-def} in the limit $\Delta \rightarrow \infty$ and respecting inequalities \eqref{ineq}.  From the definition of $\coeft_{\state^\dagger\opr^\nd \state^\nd}$ in terms of the ratio given  in  \eqref{to}, this is 
\begin{align}\label{mean-3pt}
\coeft_{\state^\dagger\opr^\nd \state^\nd} &\approx i^{-S_\opr} \bigg( \frac{c}{12} - \Delta_{\chi} - \tfrac{q_\chi^2 + \bar{q}_\chi^2}{2k} \bigg)^{3/4 - \Delta_\opr/2} \bigg( \frac{c}{12} \bigg)^{-3/4}\bigg( \Delta - \tfrac{Q^2 + \bar{Q}^2}{2k} -\frac{c}{12} \bigg)^{\Delta_\opr /2} c_{\chi^\dagger \opr ^\nd \chi} \nn  \\
 &{} \exp\bigg[ -\frac{\pi c}{3} \sqrt{ \frac{12}{c}\bigg( \Delta - \tfrac{Q^2 + \bar{Q}^2}{2k}\bigg) -1 } \bigg\{ 1 - \sqrt{ 1 - \frac{12}{c}\bigg( \Delta_{\chi} - \tfrac{q_\chi^2 + \bar{q}_\chi^2}{2k}\bigg) } \bigg\} \bigg]e^{-\frac{2\pi i }{k} ( q_\chi Q - \bar{q}_\chi\bar{Q} )}.
\end{align}
This is the central result of our work. 
The quadratic shifts by the $\uone$ charges are consistent with expectations from spectral flow.
It is also worthwhile to note that the expression above exhibits a  phase, $e^{-\frac{2\pi i }{k} ( q_\chi Q - \bar{q}_\chi \bar{Q} )}$, due to the presence of the $\uone$ charge. 


%
In the holographic (large $c$) limit, i.e, $c \gg \Delta_\chi, c\gg \Delta_O$ the above simplifies to, 
\begin{align}\label{largeCresult}
\coeft_{\state^\dagger\opr^\nd \state^\nd}  \approx N_\opr \big( \tfrac{12}{c}(\Delta - \tfrac{Q^2 + \bar{Q}^2}{2k})-1 \big)^{\Delta_O /2} c_{\chi^\dagger \opr ^\nd \chi} \, e^{ -2\pi (\Delta_{\chi} - \tfrac{q_\chi ^2 + \bar{q}_\chi^2}{2k}) \sqrt{ \frac{12}{c}( \Delta - \tfrac{Q^2 + \bar{Q}^2}{2k}) -1 } }e^{-\frac{2\pi i }{k} ( q_\chi Q - \bar{q}_\chi \bar{Q} )}.
\end{align}
Here, $N_\opr$ absorbs  the piece independent of $\Delta,Q,\bar{Q}$. 
In \S\ref{sec:einstein}, this expression will be reproduced from the $AdS_3$ dual. 

It deserves mention that one can alternatively state and derive the Cardy formula \cite{cardy1986operator} for density of states, the formula for average heavy-heavy-light coefficient \cite{Kraus:2016nwo} and the formula \eqref{mean-3pt} presented in this work, using the mathematical machinery of \textit{Tauberian theory} \cite{korevaar2013tauberian}. This line of approach has also been utilized in \cite{RychkovOPE} in the context of OPE convergence.  We refer the reader to Appendix \ref{Tauber} for a brief review of Tauberian theory, followed by a discussion of its relevance and usefulness in the present context.


\section{Asymptotics in  $\cN=2$ SCFTs}
\label{sec:n=2}
 The $\uone$ current can be naturally embedded as the R-symmetry of the $\cN=2$ superconformal algbera (see Appendix \ref{app:alg}). We shall focus on the three point coefficient of three \textit{primary} operators in $\cN=2$ SCFTs. This will enable us to provide further refinements to the asymptotic formula \eqref{mean-3pt}. {The notion of `heavy' operators is not clear for CFTs with small values of central charge and} the  analysis of this section does not include the $\cN=2$ super-Virasoro minimal models (with central charge $c=\frac{3k}{k+2}$). Since, these are the only possible $\cN=2$ SCFTs with $c\leq 3$ \cite{boucher1986determinant}, we  shall be concerned only with  SCFTs complementary to this range. 


The one-point function on the torus (of an operator $\opr$ with weights $(H,\bar H)$) can be expanded in terms of torus conformal blocks ($\cF_{ \alpha}^H(\q,y),\, \cF_{ \bar\alpha}^{\bar H}(\bar \q,\bar{y})$) as follows\footnote{Recall that the $\uone$ charge of $\opr$ is 0. Hence, it suffices to label the torus blocks by $(H,\bar H)$ which are the only non-vanishing quantum numbers.}.
\begin{align}\label{torus-block}
\ausricht
\vev{\opr} &=\sum_{i} \q^{\Delta_i - c/24} \bq^{\Delta_i -c/24} y^{q_i} \by^{q_i}\vev{i |\opr|i} \\
&= \sum_{\alpha} \vev{\alpha|\opr|\alpha} \q^{\Delta_\alpha - c/24} \bq^{\Delta_\alpha -c/24}y^{q_\alpha} \by^{q_\alpha}\cF_{ \alpha}^H(q,y) \cF_{\bar \alpha}^{\bar H}(\bq,\bar{y})  . 
\endeausricht
\end{align}
In the second equality the sum is over all primaries, labelled by $\alpha$. 
The above equation generalizes the notion of the partition function (= one-point function of the identity) which can be written as a sum of characters. For future convenience, we also rewrite the \eqref{torus-block} as an integral (similar to the previous section)
\begin{align}\label{int-rep}
\vev{\opr} = \int {\rm d}\Delta\, {\rm d}Q_+\, {\rm d}Q_-\,  T^p_{\opr}(\Delta,Q_\pm) \, e ^{ - \beta\left(\Delta-\tfrac{c}{12}\right)  } e^{2\pi i \mu_R Q_-}e^{-2\pi  \mu_I Q_+} \cF_{ \alpha}^H(\q,y) \cF_{\bar \alpha}^{\bar H}(\bq,\bar{y}) ,
\end{align}
where, the weighted spectral density for primaries, $T^p_\opr(\cdot)$, is given by  
\begin{align}
T^{p}_\opr(\Delta,Q_+,Q_-) = \sum_{i} \vev{\state_i|\opr|\state_i} \delta(\Delta-\Delta_i) \delta(Q_+ - Q_{i,+})\delta(Q_- - Q_{i,-}).
\end{align}
Here, we have resorted to the same definition of the states  $\state_i$ as in \eqref{state-def}. Since they are primaries in this context, they additionally  obey $L_n\ket{\state_i}=0$ for $n>0$.

For the one-point function of the identity operator ($H=0$), the torus blocks are simply given by the non-degenerate characters \cite{Keller:2012mr}
\begin{align}\label{susy-char}
\ausricht
\cF_{\Delta_\alpha}^0(\q,y) &= \q^{-\Delta_\alpha +{c\over 24}}y^{-q_\alpha}\chi_\alpha(\q,y) .\\
\chi_\alpha(\q,y) &= \q^{\Delta_\alpha-{c\over 24}}y^{q_\alpha}\prod_{n=1}^{\infty} {(1+y\q^{n-1/2})(1+y^{-1}\q^{n-1/2})\over (1-\q^n)^2} = \q^{\Delta_\alpha - \frac{c-3}{24}}y^{q_\alpha} \frac{\vartheta_3(\nu|\tau)}{\eta(\tau)^3}  .
\endeausricht
\end{align}
The coefficients of the $\q,y$-series count the number of descendant states and their charge at each level\footnote{Note that \eqref{susy-char} is the partition function of a theory of a complex boson and a complex fermion with central charge equaling 3, barring overall factors of $\q^{\#}$. This is the contribution from the descendants of the $\cN=2$ superconformal algebra.}. Note that the overall factor of $\q^{\Delta-c/24}y^{q_\alpha}$, which cancels out  in $\cF_{\Delta_\alpha}^0(\q,y) $,  has been accounted for in \eqref{torus-block}. 

It has been noted in \cite{Kraus:2016nwo} that in the high-energy regime $\Delta\to \infty$ and $\Delta|\log q|^2 \gg 1$, the torus block for general $H$ is dominated by the character itself.  {This has been explicitly shown to be true for Virasoro blocks in \cite{Kraus:2016nwo}; however, it is expected to be true rather generically.} That is, when the intermediate state $\ket{\alpha}$ is heavy, the insertion of the \textit{light} operator $\opr$ is a small perturbation to the degenerate character. 
\begin{align}\label{vir-1}
\cF_{\Delta_\alpha}^H(\q,y) = \cF_{\Delta_\alpha}^0(\q,y) { \left[1+  {\rm O}(\Delta_\alpha^{-1})\right]. }
\end{align}
In order to see the asymptotic  behaviour, we need to perform a S-modular transformation. Using, \eqref{susy-char} and \eqref{vir-1} and the well-known properties of the Dedekind-$\eta$ and Jacobi-theta functions we have 
\begin{align}
\ausricht
\cF_{\Delta_\alpha}^H(\q,y) 
&= \q^{1/8} \frac{(-i\tau)^{-1/2}\, e^{-\pi i \nu^2/\tau}\, \vartheta_3(\tfrac{\nu}{\tau}|\tfrac{-1}{\tau})}{(-i\tau)^{-3/2}\,\eta(-1/\tau)^3} +{\rm O}(\Delta_\alpha^{-1}),\\
&= e^{-\pi i\nu^2/\tau} (-i\tau) \q^{1/8}(\widetilde{\q})^{-1/8} +{\rm O}(\Delta_\alpha^{-1}). 
\endeausricht
\end{align}
In the final step we have retained the leading high temperature behaviour. In terms of $\beta$, the product of the holomorphic and anti-holomorphic torus blocks is\footnote{This is the analogue to equation (50) of \cite{Kraus:2016nwo}.}  
\begin{align}
\cF_{\Delta_\alpha}^H(\q,y)\cF_{\bar\Delta_\alpha}^{\bar{H}}(\bq,\by) \approx \left(\frac{\beta}{L}\right)^2 \exp\left[ -\frac{1}{4} \left( \frac{2\pi \beta}{L} - \frac{2\pi L}{\beta} \right) - \frac{  \pi   (\nu^2+\bar{\nu}^2)L}{\beta} \right].
\end{align}
 Substituting this in the one-point function \eqref{int-rep}, we have the following expression for the weighted spectral density in terms of the integral transforms (in the limit $\Delta_\alpha\to \infty$)
\begin{align}
\ausricht
&T^p_O ( \Delta, Q_\pm)   = \frac{1}{L}\bigg( \frac{2\pi}{L} \bigg)^{-\Delta_\opr} i^{S_\opr}\oint d\beta \int_{-\infty}^\infty d\mu_R\oint d\mu_I \, \,{\vev{ \opr}_{\tau, \nu} \, e^{\frac{2\pi \beta}{L} ( \Delta - \frac{c}{12} ) } e^{ -2\pi i \mu_R Q_-} e^{ 2\pi \mu_I Q_+ }\over  \cF_{\Delta_\alpha}^H(\q,y)\, \cF_{\bar\Delta_\alpha}^{\bar{H}}(\bq,\by) } .  
\endeausricht
\end{align}
Here $\vev{\opr}$ is given, as before, in terms of the dominant contribution from the lightest charged primary (and a modular transformation thereof)
\begin{align}\label{3pt}
\la \opr \ra_{\tau,\nu} &\approx \la \chi | \opr | \chi \ra |\tau|^{-2h_\opr} \  \exp \bigg\{ -  \tfrac{4\pi^2}{\beta }   \bigg(E_\chi-\tfrac{c}{12}+\tfrac{c}{6}(\nu^2+\bar{\nu}^2)+\nu q_\chi+\bar{\nu}\bar{q}_\chi\bigg)\bigg\}.
\end{align}
This leads to the same saddle equation as that of \eqref{beta-int}. However, the central charge shifts consistently all throughout the expression as 
\begin{align}\label{shifts}
c \ \mapsto \ c-3 \qquad \text{or,} \qquad k \mapsto k-1 .
\end{align}
The final expression for the Laplace transform over $\beta$ reads 
\begin{equation}
\oint d\beta\, \,  \beta^{ 1 + \Delta_\opr} \exp\bigg[ \tfrac{2\pi \beta}{L} \bigg\{ ( \Delta - \tfrac{c-3}{12} ) - \tfrac{1}{4(k-1)} ( Q_+^2 + Q_-^2 ) \bigg\} - \tfrac{2\pi L}{\beta} \bigg\{ ( \Delta_\chi - \tfrac{c-3}{12} ) - \tfrac{1}{4(k-1)} (q_+^2+q_-^2)\bigg\}\bigg].
\end{equation}
With the shifts \eqref{shifts}, the  calculation of the mean structure constant proceeds exactly in the same manner as that of the previous section. The density of states \eqref{DoS} also changes with the same shifts in the central charge. 
The refined expression for  the average value of the three-point coefficient of heavy-heavy-light primaries is then given by 
\begin{align}
\ausricht
&\coeft_{\state^\dagger O^\nd \state^\nd}^{\ p}\approx i^{-S_\opr} \bigg( \tfrac{c-3}{12} - \Delta_{\chi} - \tfrac{q_\chi ^2 + \bar{q}_\chi^2}{2(k-1)} \bigg)^{\frac{1}{4} - {\Delta_\opr\over 2}} \bigg( \tfrac{c-3}{12} \bigg)^{-3/4}\bigg( \Delta - \tfrac{Q^2 + \bar{Q}^2}{2(k-1)} -\tfrac{c-3}{12} \bigg)^{\Delta_\opr  /2} c_{\chi^\dagger\opr^\nd \chi^\nd}   \\
 &{} \exp\bigg[ -\tfrac{\pi (c-3)}{3} \sqrt{ \tfrac{12}{c-3}\bigg( \Delta - \tfrac{Q^2 + \bar{Q}^2}{2(k-1)}\bigg) -1 } \bigg\{ 1 - \sqrt{ 1 - \tfrac{12}{c-3}\bigg( \Delta_{\chi} - \tfrac{q_\chi ^2 + \bar{q}_\chi^2}{2(k-1)}\bigg) } \bigg\} \bigg]e^{-\frac{2\pi i }{k-1} ( q_\chi Q - \bar{q}_\chi \bar{Q} )}.
 \endeausricht
\end{align}
This precise shift of the central charge has also been observed in the entropy for $\cN=2$ theories \cite{Benjamin:2016aww}. The non-supersymmetric theories on the other hand show a shift of $c$ to $c-1$ \cite{Kraus:2016nwo,Cardy:2017qhl}. Although we shall not attempt to do here, it would be interesting to recover this shift from holography as well. It is tempting to speculate such a shift would be caused by the dressing of the scalars by the gravitons, gravitini and Chern-Simons $\uone$ gauge fields, thereby leading to a renormalization of the central charge.

\section{Three-point coefficients from holography}
\label{sec:einstein}
In this section we shall reproduce the 3-point coefficient, in the limit of large central charge \eqref{largeCresult} from holography. The holographic dual to the thermal state of the CFT at a non-zero chemical potential is that of a BTZ black hole, with additional $\uone$ hair (since we are interested in the energy regime above the charged BTZ threshold \eqref{ineq}). A $\uone$ Chern-Simons gauge field is also present in the bulk which is dual to the spin-1 conserved current of the CFT.  

The charged \blue{BTZ} black hole metric is exactly the same as that of the uncharged one; however, the relation between the  mass of black hole and the energy of the CFT gets modified by the non-vanishing charge of the heavy state. For the non-rotating black hole the metric is given by
\begin{align}
ds^2 = - (r^2-r_+^2 ) dt^2 + \frac{dr^2}{r^2-r_+^2} + r^2 d\phi^2 .
\end{align}
In units of  $\ell_{\rm AdS}=1$, the radius of the horizon, $r_+$ is given by,
\begin{equation}\label{rplus}
r_+  = \sqrt{\frac{12 }{c}} \left(\cL +\bar{\cL} - \frac{c}{12}  \right)^{1\over 2} =  \sqrt{\frac{12 }{c}} \left(\Delta-\frac{Q^2+\bar{Q}^2}{2k} - \frac{c}{12}  \right)^{1\over 2}.
\end{equation}
where $\cL$ and $\bar{\cL}$ are the shifted zero modes of the holographic stress tensor -- see equation \eqref{shift} of Appendix  \ref{app:u1-CS}. Note that $r_+$ is real when the condition \eqref{ineq} is satisfied.  The flat connections of the $\uone$ Chern-Simons gauge field have the following non-vanishing components
\begin{align}\label{Aw-adscft-m}
A_{w}=\frac{Q}{k}, \qquad \bar{A}_{\bar w}=\frac{\bar{Q}}{k}.
\end{align} 
Details on the derivation of the above expressions using the standard  GKPW prescription are provided in Appendix  \ref{app:u1-CS}.

 \begin{figure}[!b]
 	\centering
	\includegraphics[scale=.3]{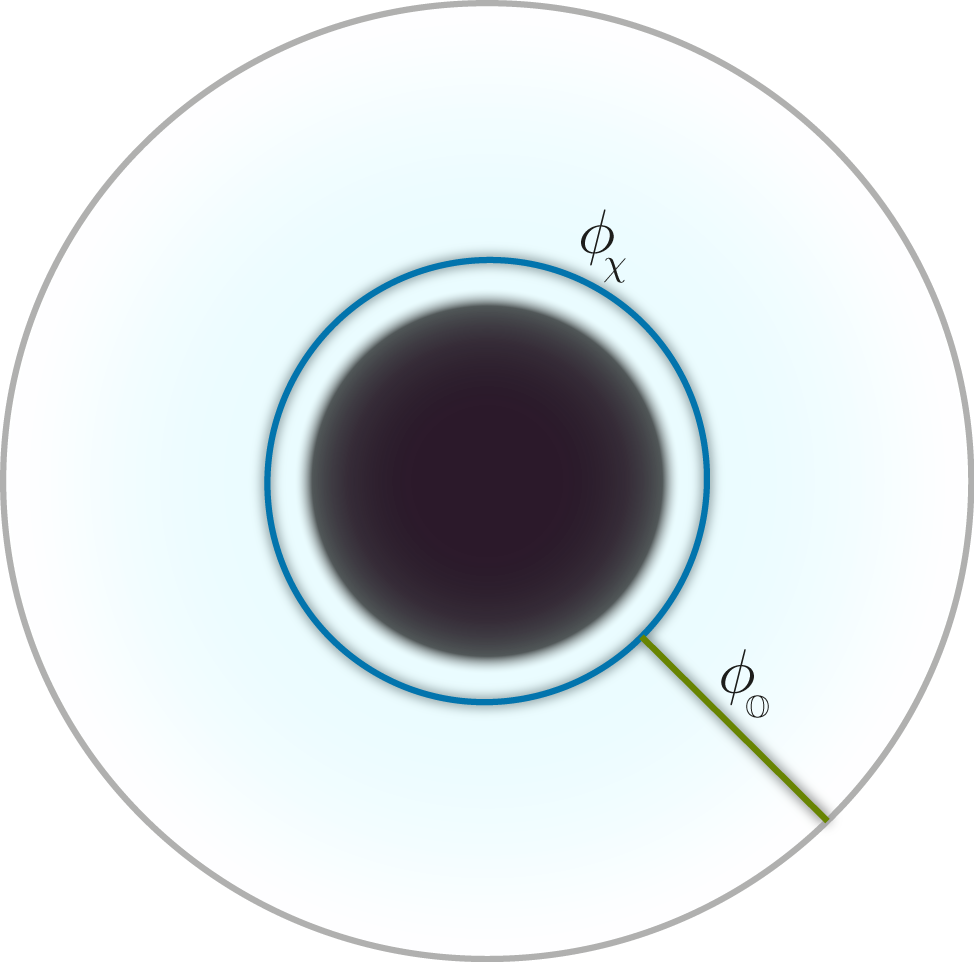}
\caption{Configuration of geodesics for the holographic calculation of the average three point coefficient in the constant time slice of the charged BTZ background. }\label{geodesics}
 \end{figure}
%
 We shall reproduce the large $c$  limit of the average three-point coefficient, $\coeft_{\state^\dagger \opr \state}$, given by equation \eqref{largeCresult}. We restrict ourselves to the regime in which the  primaries $\opr$ and $\chi$ have  $\Delta_\opr, \Delta_{\chi,q} \ll \frac{c}{12}$. 
We shall denote the bulk neutral scalar dual to the CFT probe $\opr$  by $\phi_\opr$. It has a mass $m_{\opr}$ which we take to be large. 
 Similarly $\phi_\chi$ is the charged scalar dual to $\chi$ of mass $m_\chi $, which is also large. In the large mass (bulk) limit the AdS/CFT prescription gives us,
 \begin{align}\label{masses}
\ausricht
m_\opr& \ \approx \ \Delta_\opr \ \\
m_\chi & \ \approx \ \Delta_{\chi,q} =  \Delta_\chi -\frac{q_\chi^2+\bar{q}_\chi ^2}{2k} .
\endeausricht
\end{align}
 
 We can use the geodesic approximation in the limit $1 \ll \Delta_\opr, \,  \Delta_{\chi,q} \ll \frac{c}{12}  $.  Also note that since we are interested in $\Delta_Q \rightarrow \infty$ limit, by \eqref{rplus} $r_+ \rightarrow \infty$ as well. For $\phi_\opr$, which is the bulk-to-boundary scalar, the leading amplitude is simply given in terms of its regularized length ($L \approx \log (\Lambda/r_+)$)
 \begin{align}
e^{m_\opr L} \simeq \ {\rm const.} \times r_+^{m_\opr}
 \end{align}
 This  neutral scalar $\phi_\opr$ can be thought of as arising from the fusion of two oppositely charged scalars, $\phi_\chi$ and $\phi_\chi^\dagger$, that wrap the horizon. 
 This vertex is proportional to $\vev{ \chi | \opr | \chi}$. The amplitude of the charged scalar $\phi_\chi$ wrapping the horizon in presence of the background CS field has two parts, as evident from the relevant classical Euclidean action of a charged  particle of mass $m_\chi$ and charge $(q,\bar{q})$ in the curved background:
\begin{align}\label{geo-action}
\ausricht
S^{(\chi)}_{\rm geodesic}& \ = \ S^{(\chi)}_{\rm grav} + S^{(\chi)}_{\rm CS}\ \\
S^{(\chi)}_{\rm grav}  \ = \ \ m_\chi \int d\lambda \, \sqrt{ g_{\alpha \beta} \frac{ dx^\alpha}{d\lambda}\frac{ dx^\beta}{d\lambda}}\quad  &\qquad 
S^{(\chi)}_{\rm CS}   \ =\ i q_\chi  \oint\, A_\mu dx^\mu \ - \ i \bar{q}_\chi  \oint\, \bar{A}_\mu dx^\mu  .
\endeausricht
\end{align}
Here $A_\mu, \bar{A}_\mu$ are the dual bulk Chern-Simons fields. The above integrals are along the closed loop around the horizon as shown in Fig(\ref{geodesics}). The gravitational on-shell action  simply picks up the horizon area.   
\begin{align}
S^{(\chi)}_{\rm grav}|_{\text{on-shell}} = 2 \pi m_\chi r_+ .
\end{align}

 Using the explicit solutions \eqref{Aw-adscft} for the gauge fields we can next evaluate the on-shell Wilson loop terms in $S^{(\chi)}_{\rm CS}$,
\begin{align}\label{probe-wrap}
\ausricht
S^{(\chi)}_{\rm CS }|_{\text{on-shell}} \ =   \   i q_\chi \int_{0}^{2\pi}\, A_\phi d\phi \ - \ i \bar{q} _\chi \int_{0}^{2\pi}\, \bar{A}_\phi d\phi \ = \ \frac{2\pi i }{k} \bigg( q _\chi Q- \bar{q}_\chi \bar{Q} \bigg).
 \endeausricht
 \end{align}
Putting things together, the amplitude for the geodesic configuration is 
$$
\vev{ \chi | \opr | \chi} \,  r_+^{m_\opr} e^{-2\pi r_+ m_\chi} e^{ - \frac{2\pi i}{k} ( q_\chi Q - \bar{q}_\chi \bar{Q} )}.
$$
If we use the identifications \eqref{rplus} and \eqref{masses}  we recover the large $c$ CFT result \eqref{largeCresult} for $\coeft_{\state^\dagger O^\nd \state^\nd}$, upto the overall normalization.

If $\Delta_{\chi,q}$ is of  the order of $c$, we can proceed following the arguments as presented in \cite{Kraus:2016nwo}. The massive point particle $\chi$ backreacts to give rise to a geometry with conical defect.
\begin{align}
ds^{2}=-(1+r^2)dt^2+\frac{dr^2}{1+r^2}+r^2d\phi^2, \qquad \phi \simeq \phi + 2\pi -\delta\phi
\end{align}
Here $\delta\phi$ is the deficit angle and related to mass $m_{\chi}$. Now, in charged case, $\Delta_{\chi,q}$ (rather than $E_{\chi}$, as is done for the uncharged case in \cite{Kraus:2016nwo}) has to be identified with the ADM mass, measured at infinity. Thus we obtain
\begin{align}
m_{\chi}=\frac{c}{6}\bigg(1-\sqrt{1-\frac{12\Delta_{\chi,q}}{c}}\bigg)
\end{align}
Plugging in the value of modified $m_\chi$ leads to the more general result we obtained for CFT \eqref{mean-3pt}.

The CFT result can also be reproduced using Witten diagrams, as has been done in \cite{Kraus:2016nwo}. One needs to make appropriate replacements of the conformal dimensions by their spectral flow invariant analogues. The additional phase naturally appears in the bulk-to-bulk propagator of the charged scalar field $\phi_\chi$ due to the presence of the gauge field in the bulk.

We remark that the bulk computation can also be reformulated in the Chern-Simons description of 3$d$ gravity based two copies of the gauge group  $\mathfrak{sl}(2,\mathbb{R})\times \uone$. The appropriate gauge connections for the charged BTZ black hole can be written down. Such black holes have already been constructed for the higher-spin charges (in addition to $\uone$ charge) in \cite{Datta:2013qja}. The relevant background for the present situation can be recovered by setting these higher-spin charges to zero. The 3-point function can be estimated by using a network of Wilson lines shown in Fig.~\ref{geodesics} \cite{deBoer:2013vca,Ammon:2013hba,Castro:2014mza,Hijano:2014sqa,Bhatta:2016hpz}. The Wilson loop (in the representation $\mathcal R$) corresponding to the charged scalar, $\phi_\chi$, is 
\begin{align}
\log W_{\mathcal R} (C) = \tr_{\mathcal R}(2\pi (\lambda_x -\bar\lambda_x)P_0). \nn
\end{align}
Since the scalar has mass and charge, we choose $P_0 \sim L_0 +J_0$. Here, $\lambda_x$ and $\bar\lambda_x$ are the eigenvalues of the $a_x$ and $\bar{a}_x$. (The gauge connections are of the form $A= b^{-1}a b+ b^{-1}d b$.) Evaluating the above Wilson line reproduces \eqref{probe-wrap}. The radial Wilson line attached to the boundary (of the uncharged scalar $\phi_\opr$) can be incorporated using techniques in \cite{Kraus:2017ezw}. This requires the Wilson line to be contracted with a chosen state which mimics the insertion of a primary operator in the CFT. At the trivalent vertex, the Wilson lines are joined using a suitable intertwiner, which is consistent with the CFT fusion rule \eqref{fusion}.


%
%
%
%

\def\cQ{\mathcal{Q}}

\section{Conclusions}
\label{sec:conclusions}

In this paper, we have been able to extract the mean of the heavy-heavy-light coefficient for CFTs carrying an $\uone$ chemical potential conjugate to the conserved current. Our derivation has relied on modular properties of torus one-point functions in presence of $\uone$ chemical potentials. This analysis, therefore, generalizes the one in \cite{Kraus:2016nwo}  if chemical potentials corresponding to additional conserved charges are turned on. The main result \eqref{mean-3pt} contains the same structure as that of one of the uncharged result of \cite{Kraus:2016nwo}, with conformal dimensions replaced by their spectral flow invariants. In addition, there is an extra phase present owing to the $\uone$ charge. These modifications of the uncharged result are clearly which one would have expected and it is reassuring to see these explicitly. Furthermore, the mean 3-point coefficient can also be recovered from the bulk dual. This shows that black holes with additional hairs can be suitably described as an averaged rendition of heavy microstates, which carry the additional quantum numbers corresponding to the charge. 

It would be tantalizing to understand the applicability the above results and its further generalizations. For instance, the results here can be straightforwardly generalized to the case when chemical potentials auxiliary to more spin-1 currents from a non-Abelian Kac-Moody algebra are turned on.  Another natural direction would be investigate the incarnation of the story here, if higher-spin chemical potentials are turned on. Indeed, the Cardy formula is known perturbatively in powers of the spin-3 chemical potential \cite{Gaberdiel:2012yb}. One would like to see what happens for the torus one-point function in such a case. This would also require the knowledge of the one-point functions under modular transformations. Moreover, one may hope that the average 3-point coefficient can be possibly reproduced by a suitable network of Wilson lines in the bulk \cite{Bhatta:2016hpz}. 

The formula \eqref{mean-3pt} is valid for all ranges of central charge. It will hold true for all CFTs as long as the required assumptions about the spectrum are made. These results may have a broader applicability in systems where additional $\uone$ currents are present \eg Luttinger liquids.

A major motivation behind this work arose from studying the Eigenstate Thermalization Hypothesis in CFTs. It was found in \cite{bddp} that the heavy-heavy-light coefficient plays a role in non-universal deviations of the thermal reduced density matrix and its hypothesized approximation in terms of a single heavy eigenstate. Clearly, a better understanding of ETH would require an investigation in terms of the Generalized Gibbs Ensemble (GGE), which has chemical potentials for \textit{all} conserved quantities turned on. It is therefore necessary to know the heavy-heavy-light coefficients in presence of additional chemical potentials. The result in this paper treats the simple possible case and thereby provides a small step in that direction.

In the context of holography, the three point function contains  information of bulk interaction of scalar fields. This requires bulk probes like the one considered here which are sensitive to the entire spacetime geometry. In this line of thinking, entanglement has been used as a tool to reconstruct the bulk geometry, by utilizing its   formulation in terms of the Ryu-Takayanagi surface. However entanglement entropy is a highly dynamic quantity and is susceptible to UV divergences. In contrast, the mean structure constant is not UV divergent. The infinities appearing in the lengths of the bulk geodesics which determine the three point function can thus be unambiguously removed. It would be interesting to explore this holographic relationship further and to see to what extent bulk geometry can be constrained using the average OPE coefficient.

Finally, the growth of spectral density has played a crucial role in determining whether CFTs fall within the `universality class' of being holographic. In order to admit a stringy dual, CFTs should have a sparse light spectrum and a Hagedorn growth in the density of primaries at high energies \cite{ElShowk:2011ag,Haehl:2014yla,Belin:2014fna}. For $\cN=2$ SCFTs, the properties of the elliptic genus also serve as a diagnostic for determining whether they can admit a putative gravity dual \cite{Benjamin:2015hsa,Benjamin:2015vkc}.
 It is an exciting avenue to explore the mean values of the OPE coefficient further and turn their behaviours both at the heavy and light regimes into constraints for holographic CFTs. We hope that recently developed techniques from higher genus modular bootstrap will prove to useful in this context \cite{Keller:2017iql,Cardy:2017qhl,Cho:2017fzo}.






%

\section*{Acknowledgments}

It is a pleasure to thank Alejandra Castro, Matthias Gaberdiel, Suresh Govindarajan, Shamit Kachru, Per Kraus and Alex Maloney for discussions and comments. SD thanks the   organizers and participants of the Summer Workshop at the Aspen Center for Physics, which is supported National Science Foundation grant PHY-160761. SD thanks the Department of Physics of UCSD for hospitality and an opportunity to present this work. The research of SD is supported by the NCCR SwissMap, funded by the Swiss National Science Foundation. SP thanks the organizers, participants and lecturers of TASI 2017.
 DD and SP acknowledge the support provided by the US Department of Energy (DOE) under cooperative research agreement DE-SC0009919.

\appendix 
\def\j{\mathcal{J}}
\section{Properties of Virasoro $\times\,  \uone$ algebras}
\label{app:alg}

The commutation relations of the Virasoro$\times \uone$ generators are given by  
\begin{align}
\begin{aligned}\label{comm-1}
&[L_m ,L_n] = (m-n) L_{m+n} + \tfrac{c}{12} m(m^2 -1) \delta_{m+n,0} \\
&[J_m,J_n] = \tfrac{c}{3}m\delta_{m+n,0} , \qquad \qquad [L_m,J_n] = -n J_{m+n} .
\end{aligned}
\end{align}
The OPEs involving $T$ and $J$ are
\begin{align}
\ausricht
T(z)T(w) &\sim \frac{c/2}{(z-w)^4} + {2T(w) \over (z-w)^2} + {\pd T(w) \over (z-w)} , \\
J(z)J(w) &\sim \frac{c/3}{(z-w)^2}, \\
T(z)J(w) &\sim  {J(w) \over (z-w)^2} + {\pd J(w) \over (z-w)} .
\endeausricht
\end{align}

It is well known that this algebra enjoys a spectral flow automorphism -- under this the $J_0$ and $L_0$ modes specifically transform as  ($k=c/3$)
\begin{align}\label{shift-l0}
&L'_0 =L_0 + \delta L_0 = L_0 + \eta J_0 + \frac{c}{6} \eta^2 \qquad
J'_0 = J_0 + \frac{c}{3} \eta ,
\end{align}
(see (B.21-23) of \cite{Banados:2015tft}). We can now tune $\eta$ to kill the eigenvalue of the $J_0$ charge, $\j$. This happens at $$\eta = -\frac{3}{c}\j .$$
Consequently, the $J_0$ and $L_0$ eigenvalues become
\begin{align}\label{spectral-shifts}
h' = h - \frac{3}{2c}\j^2, \qquad \j ' = 0 .
\end{align}
One can then work with these new charges for which the partition function is 
\begin{align}
Z(\tau,\nu) = \tr [\q^{L_0 -\frac{c}{24}- \frac{3}{2c}J_0^2}\, \bar \q^{\bar L_0 -\frac{c}{24}- \frac{3}{2c}\bar J_0^2}]
\end{align}
Hence, the spectrum gets shifted by the square of the charge on  both on the holomorphic and anti-holomorphic sides.

We remark that the $\cN=2$ superconformal algebra \eqref{comm-1} has additional commutation relations, given by
\begin{align}
\begin{aligned} 
&[L_m,G^{\pm}_r] = (\tfrac{m}{2}-r)G^\pm_{m+r} \quad \, \ [J_m,G^{\pm}_r] =\pm G^\pm_{m+r} \\
&\lbrace G^+_r,G^-_s  \rbrace = 2 L_{r+s} + (r-s) J_{r+s} + \tfrac{c}{3} (r^2 -\tfrac{1}{4}\delta_{r+s,0}) \qquad \lbrace G^\pm_r,G^\pm_s  \rbrace =0 . 
\end{aligned}
\end{align}
\def\cM{\mathcal{M}}
\subsection*{Unitarity bound}
\label{sec:unitarity}

We shall restrict our attention to CFTs having just Virasoro$\times\, \uone$-Kac-Moody as their chiral algebra. 
In order to obtain the unitarity constraint, we  consider the Kac determinant at level 1. For a given highest weight state $\ket{h,q}$, characterized by the conformal dimension $h$ and charge $q$, the  descendant states at level one are $L_{-1}\ket{h,q}$ and $J_{-1}\ket{h,q}$. The Gram matrix of these states is
\begin{align}
\cM^{(1)} = \begin{pmatrix}
\bra{h,q} J_1 \\ %
\bra{h,q} L_1
\end{pmatrix} \begin{pmatrix}
J_{-1}\ket{h,q} &L_{-1}\ket{h,q} 
\end{pmatrix} = \begin{pmatrix}
k   &-q \\
-q  & 2h
\end{pmatrix}.
\end{align}
In a unitary CFT, the Kac determinant should be non-negative. This gives the constraint
\begin{align}
\det [\cM^{(1)}]\,  =\, 2 k    h - q^2 \geq 0 \quad \implies \quad h - \frac{1}{2k} q^2 \geq 0,  \ \text{ for } k> 0 .
\end{align}

\section{ Chern-Simons bulk $\uone$ gauge field}\label{app:u1-CS}
The Chern-Simons action with boundary term is given by
\begin{align}\label{cs-bulk}
S_{\rm CS}=\frac{i k}{4\pi}\int_{M}\ \bigg(A\wedge dA- \bar{A}\wedge d\bar{A}\bigg) - \frac{k}{8\pi} \int_{\partial M} d^{2}x \bigg(h^{\alpha\beta}A_{\alpha}A_{\beta}+h^{\alpha\beta}\bar{A}_{\alpha}\bar{A}_{\beta}\bigg).
\end{align}
The normalisation is chosen so that $k$ appears as a level in the current algebra i.e $[J_{m},J_n]=km \delta_{m+n,0}$.
Using the AdS/CFT prescription \cite{Gubser:1998bc,Witten:1998qj}, we obtain the CFT current from the on-shell boundary variation of the bulk action \cite{Kraus:2006wn}
\begin{align}
\delta S_{\rm CS} = \frac{i}{2\pi}\int_{\partial M} d^{2}w \sqrt{h} \bigg(J^{\bar{w}}\ \delta A_{\bar{w}}-\bar{J}^{w}\ \delta \bar{A}_{w}\bigg)
\end{align}
where we use complex Euclidean coordinates $w = \phi + i t$ and $\bar{w} = \phi - it$. 
Varying \eqref{cs-bulk} and comparing with the above equation, we arrive at
\begin{align}\label{w}
J_{w}=\frac{1}{2}J^{\bar w}=\imath k A_{w}\, \qquad \bar{J}_{\bar{w}}=\frac{1}{2}\bar{J}^{w}=-\imath k \bar{A}_{\bar{w}}.
\end{align}

It is important to realize that one can either vary $A_{w}$ or $A_{\bar{w}}$, but not both. In our case, we are following the convention of \cite{Kraus:2006wn}  and varying $A_{\bar{w}}$, this leads to $J_{\bar{w}}=0$. Similarly,  for the anti-holomorphic cases, we are varying $\bar{A}_{w}$ whilst keeping $\bar{A}_{\bar{w}}$ fixed. Subsequently, from Eq.~\eqref{w}, we have
\begin{align}
J_0&=\oint dw \frac{1}{2\pi\imath}J_{w} = \oint dw \frac{k}{2\pi} A_{w} = kA_{w},\\
\bar{J}_0&=-\oint d\bar{w} \frac{1}{2\pi\imath}\bar{J}_{\bar w} = \oint d\bar{w} \frac{k}{2\pi} \bar{A}_{\bar w} = k\bar{A}_{\bar w}.
\end{align}
This implies that the components of the $\uone\times \uone$ gauge field in the bulk are 
\begin{align}\label{Aw-adscft}
A_{w}=\frac{Q}{k}, \qquad \bar{A}_{\bar w}=\frac{\bar{Q}}{k}.
\end{align}
%
%
%
The components of the stress-tensor can be obtained by variation of $h_{\mu\nu}$ of \eqref{cs-bulk}. We have
\begin{align}
T^{\text{gauge}}_{ww}=\frac{k}{4\pi}A_{w}^{2}=\frac{Q^2}{4\pi k}\\
\bar{T}^{\text{gauge}}_{\bar{w}\bar{w}}=\frac{k}{4\pi}\bar{A}_{\bar{w}}^{2}=\frac{\bar{Q}^2}{4\pi k}
\end{align}
Now, this contributes to shift in $L_0$ and $\bar{L}_0$ in following manner, leading to spectral flow invariant combinations, equation \eqref{shift-l0}.
\begin{align}\label{shift}
\delta L_{0}= - \oint\ dw T_{ww} = - \oint\ dw \frac{Q^2}{4\pi k} =-\frac{Q^2}{2k}\\
\delta\bar{L}_{0}=- \oint\ d\bar{w} \bar{T}_{\bar{w}\bar{w}} =-  \oint\ d\bar{w} \frac{\bar{Q}^2}{4\pi k} =-\frac{\bar{Q}^2}{2k}
\end{align}

\section{Applying Tauberian theorems}
\label{Tauber}
The relation between asymptotic form of a function and its Laplace transform falls naturally under the umbrella of a broad class of theorems known as \textit{Tauberian Theorems}. This branch of mathematics deals with defining infinite sums, which are otherwise not summable in the usual sense (\ie the partial sum upto $n$th term does not converge as $n \to \infty$).  In Tauberian theory, one forms a hierarchy of the notion of summability. As one goes further up the hierarchy, one can \textit{sum} {series} which are not summable in the lower hierarchy. The following example (see Chapter 1 of \cite{korevaar2013tauberian} for a detailed introduction)  elucidates the scenario.

 Consider the sum $\sum_{k=0} (-1)^{k}(k+1)$. Evidently, this is not summable in the \textit{normal sense}. But one can define
\begin{equation}\label{eq:example}
f(\beta)\equiv \sum_{k=0}(-1)^{k}(k+1)e^{-n\beta}
\end{equation}
For $\beta>0$, however, this is summable in \textit{normal sense} and we find 
\begin{align}
f(\beta)=\frac{e^{2 \beta }}{\left(e^{\beta }+1\right)^2} 
\end{align}
It's easy to see, ${e^{2 \beta }}/{\left(e^{\beta }+1\right)^2}$ is well defined at $\beta=0$ and equals to ${1}/{4}$. Thus, one can say $f(\beta)$ goes to ${1}/{4}$ as $\beta$ goes to $0$. Now one defines,
\begin{align}
\sum_{k=0}(-1)^{k}(k+1) \underbrace{=}_{\text{New notion of sum}} \frac{1}{4}.
\end{align}
This notion of sum is called \textit{Abel sum}. It's easy to see that if a series is summable in \textit{normal sense}, it is summable in \textit{Abel sense}, but not the other way around. The Tauberian theorems specify the conditions under which the higher notion of summability (\eg Abel summability) implies the lower notion of summability (\eg {normal summability}). In its generalized version, one can deduce the asymptotic behavior of \textit{normal sum} if one knows the asymptotic behavior of \textit{Abel sum}. At this point, it deserves mention that the continuous version of \textit{Abel sum} is precisely the Laplace transform.

The notion of \textit{Abel sum} (or Laplace transform as its continuous avatar)  becomes relevant, in the present context, since we are looking at quantities of the form
\begin{align}
f(\beta)=\sum_{k=0}a_{k}e^{-n\beta},
\end{align}
and trying to find out the behavior of $\sum_{k=0}a_{k}$ asymptotically. To make the analogy more precise, we note that the partition function of a CFT on a torus is given by,
\begin{align}
Z(\beta)=\int d\Delta\ \rho(\Delta)\ e^{-(\Delta-c/12)\beta}, \qquad \rho(\Delta)=\sum_{k} \delta(\Delta-\Delta_{k}),
\end{align}
where $\beta$ is the inverse temperature and length of one of the cycles of torus. Now, knowing the form of $Z(\beta)$ as $\beta$ goes to $0$ enables us to deduce the asymptotic form of $\int_{c/12}^{\Delta}\ d\Delta^{\prime} \rho(\Delta^{\prime})$ as $\Delta$ goes to $\infty$. The power of Tauberian theory comes from the fact that one does not have to impose any regularity condition on $\rho(\Delta)$.  It can be seen that deducing the Cardy formula for density of states of a CFT is equivalent to the  following theorem (for more details, see Theorem 21.1 (chapter 4) and its immediate application in Example 21.2 of \cite{korevaar2013tauberian}).
\newtheorem{theorem}{Theorem}
\newcommand\smallO{
  \mathchoice
    {{\scriptstyle\mathcal{O}}}
    {{\scriptstyle\mathcal{O}}}
    {{\scriptscriptstyle\mathcal{O}}}
    {\scalebox{.7}{$\scriptscriptstyle\mathcal{O}$}}
  }
\begin{theorem}
Let $S(v)$ be a non decreasing function and $S(v)=0$ for $v<0$. Let $F(\beta)$ be defined as
\begin{align}
F(\beta)=\int_{0}^{\infty}\ e^{-\beta v}\ dS(v)=\beta \int_{0}^{\infty}\ S(v) e^{-\beta v} dv, \qquad \text{for }\ \text{Re}(\beta) >0,
\end{align}
and
\begin{align}
e^{-f(\beta)}F(\beta) \to 1, \qquad \text{as }\ \beta\to 0
\end{align}
uniformly in every angle $|\text{arg}(\beta)| \leq\beta_{0}<\frac{\pi}{2}$. Furthermore, if $f(\xi)$ satisfies the following conditions as $\xi \in \mathbb{R}$ and $\xi \searrow 0$, in which we have $0<\delta(\xi)\leq \frac{\xi}{2}$, 
\begin{itemize}
\item $f(\xi)$ is real and positive,
\item $-\xi f^\prime(\xi)\nearrow \infty$,
\item $\frac{\sqrt{f^{\prime\prime}(\xi)}}{|f^\prime(\xi)|}=\smallO \left(\frac{\delta(\xi)}{\xi}\right)$,
\item $f^{\prime\prime}(\xi+z)= \mathcal{O}\left(f^{\prime\prime}(\xi)\right)$ uniformly for $|z|\leq \delta(\xi)$.
\end{itemize}
Then, 
\begin{align}
S(v) \sim S_0(v)=\frac{e^{vh(v)+f(h(v))}}{h(v)\sqrt{2\pi f^{\prime\prime}(h(v))}}, \qquad \text{as } v \to \infty
 \end{align}
\end{theorem}

To derive the Cardy formula \cite{cardy1986operator}, we make following identifications\footnote{{In terms of $\Delta$, the theorem requires that $F(\beta)=\int_{c/12}^{\infty}\ d\Delta\ \rho(\Delta) e^{-\beta(\Delta-c/12)}$. The partition function,  $Z(\beta)=\int_{0}^{\infty}\ d\Delta\ \rho(\Delta) e^{-\beta(\Delta-c/12)})$, however differs from $F(\beta)$. Nonetheless, in $\beta\to0$ limit, the dominant contribution to $Z(\beta)$ arises from large $\beta$. Hence, the fact $Z(\beta)$ differs from $F(\beta)$, does not affect this result. In the $\beta\to0$ regime, we indeed have $F(\beta)=Z(\beta)$.} }
\begin{align}
F(\beta)=Z(\beta),\qquad 
f(\beta) =\frac{\pi^{2}c}{3\beta},\qquad 
v \to \Delta-\frac{c}{12},\qquad 
S(v) \to \int_{c/12}^{\Delta}\ d\Delta^{\prime}\ \rho(\Delta^{\prime}),
\end{align} 
and concretely, this yields
\begin{align}
\int_{c/12}^{\Delta}\ d\Delta^{\prime}\ \rho(\Delta^{\prime}) \underset{\Delta \to \infty}{\sim} \frac{1}{2\sqrt{\pi}}\left(\frac{\pi^2c}{3}\right)^{-1/4}\bigg(\Delta-\frac{c}{12}\bigg)^{-1/4}\exp\bigg(4\pi\sqrt{\frac{c}{12}(\Delta-\frac{c}{12}})\bigg).
\end{align}

Now, one can argue that heavy excited states lie very closely to each other so that density of states becomes almost continuous, which allows us to take a derivative with respect to $\Delta$ and deduce the Cardy formula \cite{cardy1986operator} for density of states of a CFT 
\begin{align}
\rho(\Delta) \underset{\Delta \to \infty}{\sim} \frac{1}{\sqrt{2}}\left(\frac{c}{12}\right)^{1/4}\bigg(\Delta-\frac{c}{12}\bigg)^{-3/4}\exp\bigg(4\pi\sqrt{\tfrac{c}{12}(\Delta-\tfrac{c}{12}})\bigg)
\end{align}

 Similarly, one can justify the result of \cite{Kraus:2016nwo} by appealing to same theorem provided the quantity estimated \ie  $\int_{c/12}^{\Delta} d\Delta^{\prime}\ T_{\opr}(\Delta^{\prime})$ is a non decreasing function of $\Delta$. This requirement can be however relaxed by observing that if all the $c_{\state^\dagger\opr^\nd\state^\nd}$ comes with same phase, one can then define a new quantity $\tilde{T}_{\opr}$ by absorbing the phase. This ensures that $\int_{c/12}^{\Delta}\ d\Delta^{\prime}\ \tilde{T}_{\opr}(\Delta^{\prime})$ is a non decreasing function of $\Delta$. The more general scenario without this caveat can be explored by using the Tauberian theory of more general function $S(v)$. For now, with this caveat in mind, $\int_{c/12}^{\Delta} d\Delta^{\prime}\ T_{\opr}(\Delta^{\prime})$ and $\int_{c/12}^{\Delta} d\Delta^{\prime}\ \rho(\Delta^{\prime})$ are obtained.
 Once again, going from integral to integrand requires taking derivative and is justified by appealing to the extra input that at heavy excited states $T_{\opr}$ becomes a smooth function of $\Delta$. Hence, this justifies the definition
 \begin{align}\label{KMjust}
\coeft_{\state^\dagger\opr^\nd \state^\nd} =  \frac{T_{\opr}(\Delta)}{\rho(\Delta)}.
\end{align}
It is worthwhile to note that one can define following cumulative average as well
\begin{align}\label{KMjust1}
\coeft_{\state^\dagger\opr^\nd \state^\nd}^{\text{c-avg}} = \frac{\int_{c/12}^{\Delta} d\Delta^{\prime}\ T_{\opr}(\Delta^{\prime})}{\int_{c/12}^{\Delta} d\Delta^{\prime}\ \rho(\Delta^{\prime})}.
\end{align}
In the large $\Delta$ limit, the above quantity has exactly the same exponential behavior as $\coeft_{\state^\dagger\opr^\nd \state^\nd}$.  

It deserves mention that  in order to apply the above theorem for the OPE coefficient, one needs to ensure $f(\beta)$ is real and positive in the limit $\beta \searrow0$. Since, $f(\beta)$ goes like $-\Delta_{\opr}\log(\beta)+\frac{4\pi^{2}}{\beta}\left(\frac{c}{12}-\Delta_{\chi}\right)$, the positivity is guaranteed assuming existence of state $\chi$ such that $\Delta_{\chi}< \frac{c}{12}$. (Note that, $\log(\beta)< 0$ for $\beta<1$ and $\Delta_{\opr}>0$ by unitarity bound.)  In fact, $\Delta_{\chi}< \frac{c}{12}$ becomes a necessary condition as $-\beta f^{\prime}(\beta) >0 $ as $\beta \searrow 0$ is required for the above theorem to hold. Furthermore,  one can extend the prescription and use the Tauberian machinery to the charged case which we have studied here. In the charged case, where $\chi$ carries charge $(q,\bar{q})$, we have $\Delta_{\chi}$ replaced by $\Delta_{\chi,q}=\Delta_{\chi}-\tfrac{1}{2k}\left(q_{\chi}^2+\bar{q}_{\chi}^2\right)$ and the requirement becomes $\Delta_{\chi,q}<\frac{c}{12}$.

\begin{small}
	\bibliography{collection}
	\bibliographystyle{bibstyle}
\end{small}
\end{document}